\documentclass[a4paper, amsfonts, amssymb, amsmath, reprint, showkeys, nofootinbib, twoside]{revtex4-1}

\usepackage[compat=1.0.0]{tikz-feynman}
\usepackage{subfigure}

\usepackage[english]{babel}
\usepackage[utf8]{inputenc}
\usepackage{gensymb}
\usepackage[colorinlistoftodos, color=green!40, prependcaption]{todonotes}
\usepackage[pdftex, pdftitle={Article}, pdfauthor={Author}]{hyperref} 
\begin{document}

\title{Tau Neutrinos with  Cherenkov Telescope Array}
\author{Damiano Fiorillo, Gennaro Miele, Ofelia Pisanti}
    \affiliation{Dipartimento di Fisica {\it "Ettore Pancini"}, Universit\`a degli studi di Napoli Federico II and INFN - Sezione di Napoli, Complesso Univ. Monte S. Angelo, I-80126 Napoli, Italy}

\date{\today} 

\begin{abstract}
The next generation of Imaging Atmospheric Cherenkov telescope, like CTA, is going to strongly improve the detection capability of high-energy cosmic rays. In our paper we discuss the possibility to use such apparatus to detect Earth-skimming tau neutrinos. Interestingly the analysis shows that order few events per year can be detected for energies above $10^8$ GeV in the optimistic case of larger neutrino fluxes  produced by Flat Spectrum Radio Quasars. However, even for more conservative cosmogenic neutrino fluxes such rate will be also obtained, but for a decade of running. This estimate seems to candidate a set up like CTA for performing high energy neutrino astronomy as well.
\end{abstract} 
\maketitle

\section{Introduction}

Ultra High Energy neutrinos (UHE$\nu$) provide a unique probe for new physical interactions and to unveil the mechanisms at work in  extreme astrophysical environments. The quite recent observations of astrophysical neutrinos in the TeV-PeV energy range by Neutrino Telescopes have further stimulated the interest of the scientific community about the so-called neutrino astronomy. 

Neutrinos with such extreme energy, namely above $10^{17}$-$10^{18}$ eV, are expected to be produced from the interaction of UHE cosmic rays with the Cosmic Microwave Background (CMB) {\it via} the $\pi$-photoproduction, $p + \gamma_{CMB} \rightarrow n + \pi^+$ during their propagation in the universe, the so-called {\it cosmogenic neutrinos} \cite{Beresinsky:1969qj} or {\it GZK-neutrinos} and they have been extensively studied in number of papers, see for instance Ref.s \cite{stecker1973ultrahigh,berezinsky1975cosmic,hill1983ultra,engel2001neutrinos,kalashev2002ultrahigh,semikoz2004ultra,allard2006cosmogenic,Anchordoqui:2007fi,takami2009cosmogenic,Berezinsky:2010xa,Gelmini:2011kg,gelmini2012gamma,stanev2014cosmogenic,roulet2013pev,aloisio2015cosmogenic,Heinze:2015hhp,Halzen:2016gng}. However, the prediction for such a flux is still affected by severe uncertainties, mainly concerning the spatial distribution of astrophysical sources for the cosmic rays and their nature, the precise form and the chemical composition of ejected hadron fluxes (if proton or different nuclei), and the way of modelling the diffuse extragalactic electromagnetic background in the different frequency regions. Furthermore alternative sources of UHE$\nu$ have been proposed, where neutrinos are straightforwardly produced at extreme astrophysical sources as a secondary product of the acceleration of hadronic matter, see for instance  Ref. \cite{Murase:2014foa}.

\begin{figure}[htbp]
\begin{center}
\includegraphics[width=0.8\columnwidth]{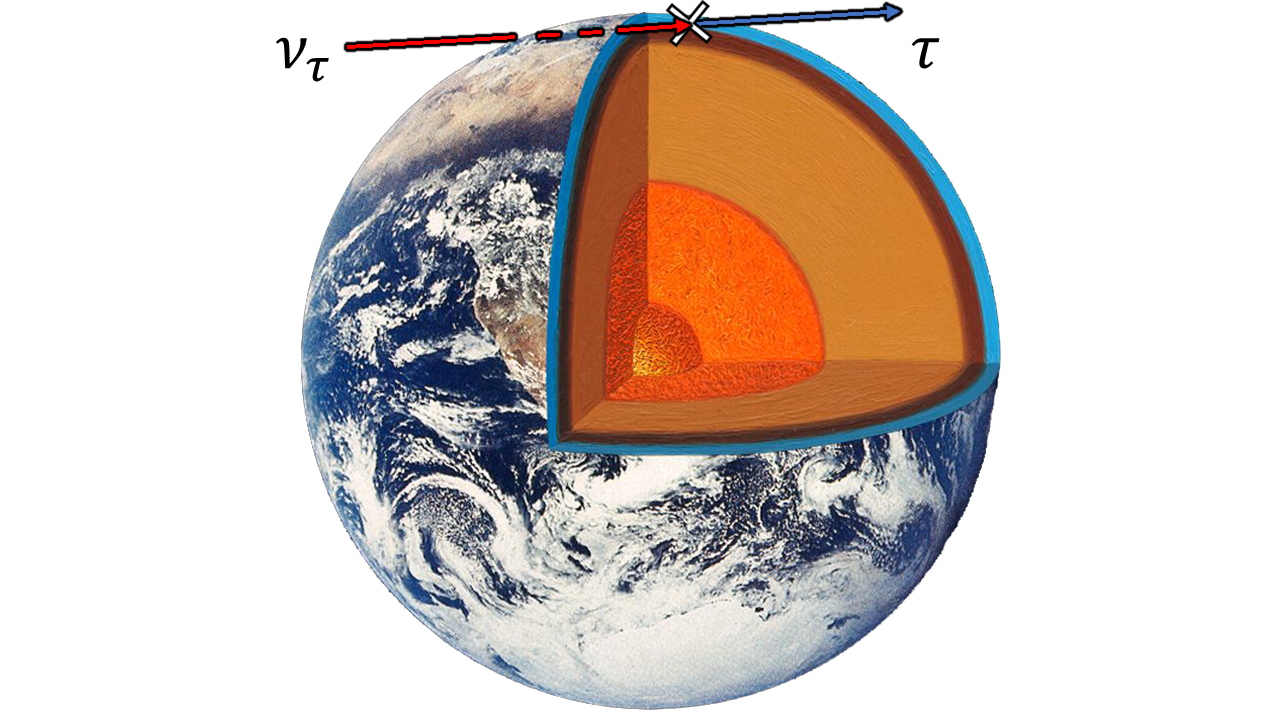}\caption{
A $\nu_\tau$ Earth-skimming event.}\label{earthskimming}
\end{center}
\end{figure}

Unfortunately, UHE neutrinos can be hardly observed if compared with other standard particles. The interaction length of an EeV neutrino is about 500 km water equivalent in rock and, and even crossing horizontally the atmosphere (360 meters water equivalent), only one neutrino out of thousand will be interacting. The small neutrino-nucleon cross section and a very low expected flux impose to use for their detection giant apparatus, like  km$^3$-Neutrino Telescopes (IceCube \cite{Aartsen:2013jdh,Aartsen:2014gkd}, Antares \cite{Collaboration:2011nsa}, Km3Net \cite{Adrian-Martinez:2016fdl}) or cosmic rays set up's, like  Pierre Auger Observatory \cite{Abraham:2004dt, Abraham:2008ru} or planned ones, like IceCube-Gen2 \cite{vanSanten:2017chb}, Auger-Prime \cite{Martello:2017pch}, ARCA \cite{Piattelli:2015pmp}, ORCA \cite{Brunner:2015ltd}, BAIKAL-GVD \cite{Avrorin:2013uyc}, GRAND \cite{Alvarez-Muniz:2018bhp} and JEM-EUSO \cite{Adams:2013vea}. However, an interesting strategy can be applied for $\nu_\tau$ detection as described in Ref.s \cite{Capelle:1998zz,Halzen:1998be,Fargion:1999se,Fargion:2000iz,Becattini:2000fj,Dutta:2000hh,Beacom:2001xn,Dutta:2002zc,Bottai:2002nn,Kusenko:2001gj,Feng:2001ue,Bertou:2001vm,Aramo:2004pr,Miele:2005bt,Cuoco:2006qd}. Interestingly, for energy between $10^{18}$ and $10^{21}$ eV the $\tau$ decay-length is of the order of the corresponding interaction range. Hence, an energetic $\tau$, produced by a $\nu_\tau$ not too deep under the surface of the Earth, has a real chance to emerge in the atmosphere as an up-going particle. and once decayed to produce a shower. A muon with the same energy would instead loose its energy too fast hence stopping and decaying in the rock. Thus considering almost horizontal $\nu_\tau$, just skimming the Earth surface (typically denoted Earth-skimming $\nu_\tau$), they would cross an amount of rock of the order of their interaction length and thus emerging from the surface could produce a shower potentially detectable (see Fig. \ref{earthskimming}).
In these years, parallel to the development of Neutrino Astronomy, it has been growing the use of ground-based gamma-ray detectors.  This is a quite novel field of research with enormous capability to study new physics. Since the first detection in 1989 of a signal at TeV from the Crab nebula, with the Whipple 10m Imaging Atmospheric Cherenkov Telescope (IACT), developments of new techniques have settled down the astronomy with IACTs that now counts with major arrays like  H.E.S.S. \cite{Hinton:2004eu}, MAGIC \cite{Mirzoian:2004bk}, and VERITAS \cite{Holder:2006gi}. The Cherenkov Telescope Array (CTA) \cite{Consortium:2010bc} represents the next generation of IACTs and, with two sites placed in the two hemispheres, will allow the observation of the whole sky with high sensitivity and angular resolution above a few tens of GeV. CTA is supposed to improve our understanding of the high-energy phenomena occurring in the universe and will allow us to explore fundamental open questions in physics.

In the present paper we investigate the possibility to use an apparatus like CTA for detecting Earth-skimming UHE$\nu_\tau$. To this aim, in Section \ref{sec:effective} we introduce the concept of effective detection area, whereas Section  \ref{sec:flux}  is devoted to the computation of up-going $\tau$-flux. In Section \ref{sec:simulation} and \ref{sec:manytel} the detection efficiency simulation for up-going $\tau$ is described. Section \ref{sec:results} reports our results, followed by our conclusions and remarks.
\section{Effective areas}
\label{sec:effective}

Let us denote by  $\eta(E_\tau, \Omega, x, y)$ the efficiency of the considered IACT in the detection of a shower induced by a $\tau$ with energy $E_\tau$, direction of exit from the Earth, $\Omega$ (up-going), and exiting point of coordinates $(x, y)$\footnote{We are here neglecting the curvature of the Earth; the condition for doing so is that the distances at which the shower is produced are much smaller than the Earth radius. We discuss in Appendix \ref{app2} how this assumption should be modified in case the distances should become too large.}. Then the total number of detected events is
\begin{equation} \label{numev}
\frac{dN_\tau}{dt}=\int dE_\tau\,  d\Omega \int dx \, dy \frac{d\Phi_\tau (E_\tau, \Omega)}{dE_\tau \, d\Omega}  \eta(E_\tau,\Omega, x, y).
\end{equation}
Here $d\Phi_\tau/(dE_\tau d\Omega)$ represents the tau-lepton flux exiting the Earth. As shown in Section \ref{sec:simulation} the quantity $\eta(E_\tau, \Omega, x, y)$ has to be obtained by a Monte Carlo simulation.\\
Due to homogeneity and isotropy of the incident neutrino flux,  $d\Phi_\tau/(dE_\tau d\Omega)$ cannot depend on the point $(x,y)$, hence $\eta$ can be integrated over the coordinates $x$ and $y$, yielding the $\tau$ effective area, $A_{eff}^\tau (E_\tau,\Omega)$, and the Eq. \eqref{numev} takes the form
\begin{equation}
\frac{dN_\tau}{dt}=\int dE_\tau\, d\Omega \frac{d\Phi_\tau (E_\tau, \Omega)}{dE_\tau d\Omega}  A_{eff}^\tau (E_\tau,\Omega).
\label{eq:Aefftau}
\end{equation}

Since the tau-leptons are produced \textit{via} Charged Current (CC) interactions, the linearity of the transport processes of neutrinos through the Earth allows us to write the $\tau$ flux as
\begin{equation}
\frac{d\Phi_\tau(E_\tau, \Omega)}{dE_\tau d\Omega}  =\int dE_\nu\, d\Omega' \frac{d\Phi_\nu(E_\nu, \Omega')}{dE_\nu d\Omega'} k (E_\nu,E_\tau,\Omega',\Omega).
\label{eq:k}
\end{equation}
Due to the large energy of the CC-event one can safely assume $k (E_\nu,E_\tau,\Omega',\Omega)=K(E_\nu,E_\tau,\Omega)\delta(\Omega'-\Omega)$; hence one gets 
\begin{eqnarray}
\frac{dN_\tau}{dt}=\int dE_\tau\, dE_\nu\, d\Omega \frac{d\Phi_\nu (E_\nu, \Omega)}{dE_\nu d\Omega}  A_{eff}^\tau (E_\tau,\Omega) K(E_\nu,E_\tau,\Omega) \nonumber\\
\equiv \int dE_\nu \, d\Omega \frac{d\Phi_\nu (E_\nu, \Omega)}{dE_\nu d\Omega}  A_{eff}^\nu (E_\nu,\Omega),\,\,\,\,\,\,\,\,\,\,\,\,\,\,\,\,
\label{eq:Aeffnu}
\end{eqnarray}
where the integral kernel, $K(E_\nu,E_\tau,\Omega)$, has to be calculated taking into account the processes that produce the up-going $\tau$ flux. The quantity $A_{eff}^\nu (E_\nu,\Omega)$ hence denotes the $\nu_\tau$ effective area of the apparatus and it is defined as
\begin{equation}
A_{eff}^\nu (E_\nu,\Omega) \equiv \int  dE_\tau \, A_{eff}^\tau (E_\tau,\Omega) K(E_\nu,E_\tau,\Omega) \,\,\,.
\end{equation}
In the following we will clarify how to compute the integrand of previous equation.

\section{The flux of up-going tau-leptons}
\label{sec:flux}

In this section we calculate the flux of up-going tau-leptons that enters Eq. (\ref{eq:Aefftau}). For a given spectrum of tau neutrinos entering the Earth, the evolution of the tau flux inside the Earth is regulated by three key processes: the production of tau leptons from CC interactions of tau neutrinos, the decay of tau leptons and their energy losses.

Let us consider a certain line of propagation for a $\tau$, defined by the direction $\Omega$ in the laboratory rest-frame, and let $x$ be the equivalent thickness in Earth, defined by $dx=\rho(l) dl$. By denoting with $T(E,x)$ the flux of $\tau$ of given energy $E$ and at a given thickness $x$, and with $N(E',x)$ the flux of $\nu_\tau$, the evolution is regulated by the transport equation
\begin{eqnarray} \label{propaeq}
\frac{\partial T}{\partial x}-\frac{\partial}{\partial E} \left( B(E) T\right)&=&\int_{E}^{+\infty} dE' \frac{N(E',x)}{m_a} \frac{d\sigma_{CC}(E'\rightarrow E)}{dE}\nonumber\\
&-&\frac{m_\tau T}{c\tau E\rho(x) }.
\end{eqnarray}
Here $B(E)=E(\alpha + \beta E^s)$ describes the tau energy losses, and the last term corresponds to the tau-lepton decay, where the factor $m_\tau/E$ describes the relativistic time dilation. This equation can be rearranged as
\begin{eqnarray}
&&\frac{\partial T}{\partial x}-B(E) \frac{\partial T}{\partial E} =T \frac{\partial B}{\partial E}\nonumber\\
&+&\int_{E}^{+\infty} dE' \frac{N(E',x)}{m_a} \frac{d\sigma_{CC}(E'\rightarrow E)}{dE}-\frac{m_\tau T}{c\tau E \rho(x) },
\end{eqnarray}
and it admits an analytical solution with the method of characteristics. If $x$ is measured from the exit point of the tau-lepton backwards, and denoting its final energy by $E$, the energy at a position $x$, $\bar{E}(x)$, is defined by the equation
\begin{equation}
\int_E^{\bar{E}(x)} \frac{d\epsilon}{B(\epsilon)}=x,
\end{equation}
whose solution is
\begin{equation}
\bar{E}(x)=E e^{\alpha x}\left[\frac{\alpha}{\alpha+\beta E^s (1-e^{\alpha s x})}\right]^{1/s}.
\end{equation}
Then the tau-lepton flux at the exit of the Earth is
\begin{eqnarray}
T(E,X)=\int_0^X dx \exp\left\{-\int_0^x dx'\left[\frac{m_\tau}{c\tau \bar{E}(x') \rho(x') } \right. \right. \,\,\,\,\,\\
\left. \left. -\frac{\partial B(\bar{E}(x'))}{\partial E}\right]\right\} \int_{\bar{E}(x)}^{+\infty} \frac{N(E',x)}{m_a} \frac{d\sigma_{CC}(E'\rightarrow \bar{E}(x))}{d\bar{E}(x)} dE'.\nonumber
\end{eqnarray}
Here $X=X(\Omega)$ is the total thickness traversed by the neutrinos from the entering to the exiting of the Earth, that is the thickness corresponding to the geometrical length traversed by the neutrinos, $L(\Omega)$.

At high energies it has been shown that the results obtained are reasonably approximated by writing the cross section as
\begin{equation}
\frac{d\sigma_{CC} (E'\rightarrow E)}{dE}=\sigma_{CC}(E') \delta(E-(1-y) E'),
\end{equation}
where $y$ is the mean inelasticity, around $0.2$. Some further manipulations lead to the result, now expressed as a function of the angle
\begin{eqnarray}
&&T(E,\Omega)=\frac{1}{m_a(1-y)} \int_0^{L(\Omega)} dl \frac{B[\bar{E}(x(l))]}{B(E)} \rho(l) \theta\left[x_c-x(l)\right] \nonumber\\
&\times & \exp\left[-\int_0^l \frac{m_\tau dl'}{c\tau \bar{E}(x(l'))}\right]\sigma_{CC} \left[\frac{\bar{E}(x(l))}{1-y}\right]N\left[\frac{E(x(l))}{1-y},x(l)\right], \nonumber\\
\label{solu}\end{eqnarray}
where $x(l)=\int_0^l \rho(l') dl'$, while the value $x_c$ in the Heaviside theta,
\begin{equation}
x_c=\frac{1}{\alpha s} \log\left[1+\frac{\alpha}{\beta E^s}\right],
\end{equation}
is the value of $x$ for which $E(x)$ diverges, that is, the maximum length from which a tau can come without losing too much energy.

$T(E,\Omega)$ depends on the flux of $\nu_\tau$, $N(E',x)$, which should be determined of course by solving explicitly the coupled differential equation. We make instead the conservative assumption that regeneration, both by tau-leptons decay and by neutral current scattering processes, can be neglected: therefore, our method underestimates the fluxes of particles produced. However, this is not expected to make a too large difference, since the neutrinos relevant for the detection are skimming, and therefore only traverse a short path inside the Earth, making regeneration weak. Under this hypothesis, we find
\begin{eqnarray}
&&N\left(E,x\right)=N\left[E,X(\Omega)\right]\nonumber\\
&\times & \exp\left[-(\sigma_{CC} (E) + \sigma_{NC} (E)) \frac{X(\Omega)-x}{m_a}\right].
\end{eqnarray}
The information on the propagation is completely contained in the integral kernel of \eqref{propaeq}. This is obtained from \eqref{solu} by choosing a monochromatic neutrino spectrum $N(E)=\delta(E-E_\nu)$. We thus find that the integral kernel, $K(E_\nu, E_\tau)$, is
\begin{eqnarray}
K(E_\nu,E_\tau,\Omega)=\frac{B\left[E_\nu(1-y)\right]}{B[E_\tau]} \frac{1}{m_a \frac{dE}{dx}} \sigma_{CC} (E_\nu) \nonumber\\
\times  \exp\left[-\int_0^{l(E_\nu,E_\tau)} \frac{m_\tau dl'}{c\tau \bar{E}(x(l'))}\right]\nonumber\\. \times \exp\left[-(\sigma_{CC} (E_\nu) + \sigma_{NC} (E_\nu)) \frac{X(\Omega)-x(l(E_\nu,E_\tau))}{m_a}\right],\nonumber\\
\end{eqnarray}
where $l(E_\nu,E_\tau)$ is the length at which the tau is produced, such that
\begin{equation}
\bar{E}(x(l))=E_\nu (1-y).
\end{equation}
Notice that the kernel depends implicitly upon $\Omega$.

\section{Detector efficiency simulation}
\label{sec:simulation}

\begin{figure}[h!]
\begin{center}
\includegraphics[width=.6\columnwidth]{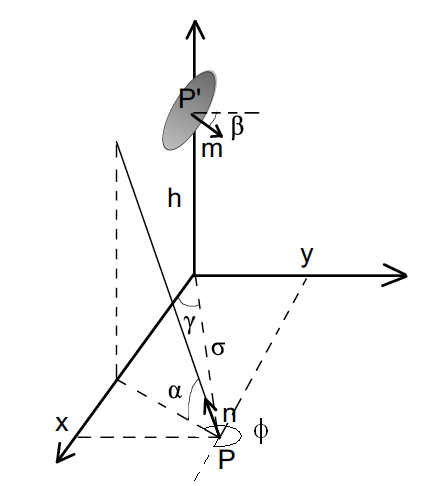}\caption{Geometrical quantities used in the simulation. For the clarity of the picture the detector is placed at a certain altitude  $h$ from the ground, even though such quantity is neglected in the calculations.}\label{geometry1}
\end{center}
\end{figure}

In order to perform a  numerical  simulation we have to fix the geometry of the apparatus. Let us model each detector as a circular plane telescope disposed on the ground, whose normal, $\mathbf{m}=(0,\cos\beta,-\sin\beta)$, is placed for simplicity in the $y-z$ plane (see Figure \ref{geometry1}). The slight altitude, $h$, at which the telescope mirror is set will be neglected in our calculations.
Let us denote with ${P=(x,y)}$ the point on the ground from which the $\tau$ is emerging with direction $\mathbf{n}=(\cos\alpha\,\cos\phi,\cos\alpha\,\sin\phi,\sin\alpha)$. Using polar coordinates in the x-y plane  (see Figure \ref{geometry1}) we have $x=\sigma\cos\gamma$, $y=\sigma\sin\gamma$.  Note that an effective detection corresponds to having $\phi\sim\gamma + \pi$.

A necessary ingredient for the simulation is the number of Cherenkov photons produced by the shower and intercepting the telescope. The detection will generally depend on whether or not the number of these photons is higher than a threshold value. A great simplification results from the natural assumption that the dimension of each detector, which ranges from few meters up to 23 m of diameter, is much smaller than the typical transverse size of the showers.

\begin{figure}[h!]
\begin{center}
\includegraphics[width=.7\columnwidth]{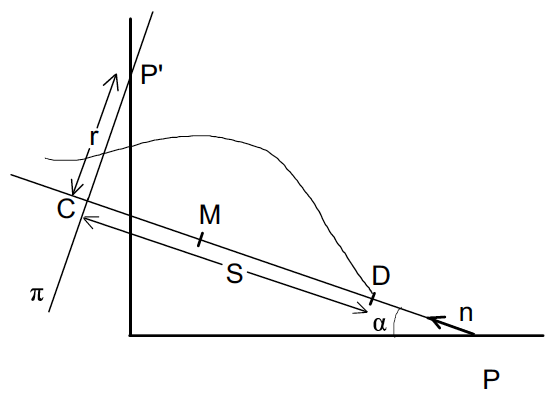}\caption{Cherenkov production from the $\tau$ shower.}\label{geometry2}
\end{center}
\end{figure}

In order to check this assumption, let us consider the plane, $\pi$, orthogonal to the direction of the shower, $\mathbf{n}$, that contains the center of the detector, $P'$ (see  Figure \ref{geometry2}, where we consider the simple case of a shower coming from the positive $y$ axis) and let us denote by $C$ the shower core on $\pi$. The Cherenkov cone arriving on the detector has to be at least as large as $\theta S$, where $\theta$ is the Cherenkov angle and $S$ stands for the distance along the shower axis from the $\tau$ decay point, $D$, to $C$. The latter quantity is typically larger than $10$ km, hence the lateral profile has a size larger than $\sim 100$ m. A simple numerical analysis confirms the validity of this approximation. Having ascertained this, we can approximate the number of photons intercepting the detector as simply the Cherenkov photon density at the point $P'$ times its area. 

The number of photons crossing the unit area perpendicular to the shower axis at the point $P'$, hereafter denoted by $dQ/dA$, is a function of $S$, and of $r$, namely the distance between $C$ and $P'$. The computation of $dQ/dA$ is reported in Appendix \ref{app1}. In general, one has to take into account the fact that the $\tau$-lepton does not decay immediately at high energies. In fact, we should in principle average our results over the decay distribution of the particle. Here we adopt the simplification that the tau decays exactly in the point D at the distance, $l_\tau(E_\tau)$, corresponding to the mean value of the distribution, that is, after a time equal to the lifetime of the lepton in the tau comoving frame. Of course, if the position of D were beyond the plane $\pi$, then the shower should be regarded as not detected. Moreover, another comment is in turn. While it is true that Cherenkov light is emitted by each portion of the shower, at the same time most of this light concretely comes from the shower maximum, M. If the point M lies beyond the plane $\pi$, then the cone which connects it to the detector surface does not intersect the front surface, but rather the back surface. In this condition, one can reasonably expect that the light reaching the detector results in a negligible signal.

All these geometrical constraints have been implemented in our simulation. In particular, for a given $\tau$ energy, $E_\tau$, and shower direction, $\Omega=(\alpha,\phi)$, we vary the angle $\gamma$ and determine the range of $\sigma$ for which the number of photons arriving to the detector is larger than the detection threshold. The spot at ground resulting from the set of all the corresponding points $P$, obtained when $\gamma$ is varied, gives the $\tau$ effective area, $A_{eff}^\tau(E_\tau,\Omega)$ of Eq. \ref{eq:Aefftau}. This calculation is performed for a grid of  $E_\tau$ and $\Omega$.

We report in Appendix \ref{app2} the values of the geometrical quantities defined in this section and the constraints implemented in the simulation. We also provide there the corrections that one has to implement to take into account the Earth curvature.

\section{Generalization to many telescopes}
\label{sec:manytel}

An important feature of the CTA is connected with the presence of many telescopes, fixed at distances between one another of the order of the hundreds of meters. As we mentioned in the previous section, for a single telescope tau lepton detection will be efficient for leptons coming from a definite geometrical region on the Earth surface. The question naturally arises whether it is possible to improve the sensitivity to the skimming neutrinos by suitably orienting the telescopes to detect a larger number of events, so that the area of this geometrical region increases. Since the typical distances over which leptons travel in order to be detected from a single telescope are of the order of tens to hundreds of kilometers, which is much larger than the distance between the telescopes, we expect that if different telescopes all look in the same direction, they will observe tau leptons coming from the same region, thereby not improving upon the single telescope performance. However, if the direction in which the telescopes are looking is different, they will be able to probe different regions and therefore enlarging the effective area of the resulting system. It is therefore expected that the optimum configuration for the detection of tau leptons is obtained by orienting as many telescopes as possible in different angular directions, so as to cover the entire $2\pi$ angle. \\ It is clear that, if the detectable region for a single telescope has a finite angular width $\delta\gamma$, the largest number of telescopes which can be oriented in different angular directions without their detectable regions intersecting will be $\frac{2\pi}{\delta\gamma}$. If this number is smaller than the total number of available telescopes, then the largest effective area which can be attained will be
\begin{equation}
A_{eff}^{max}=\frac{2\pi}{\delta\gamma} A_{eff}.
\end{equation}
On the other hand, if this number is larger than the total number of available telescopes, the latter will be the saturation number. \\ By this method we are able to generalize the single telescope effective area to a many telescope effective area. We have used this methodology separately for the three types of telescopes at CTA, the Small Sized Telescopes (SST), the Medium Sized Telescopes (MST) and the Large Sized Telescopes (LST). The total number of telescopes of each class at the CTA South Site is respectively $70$, $40$, and $4$ \cite{Knodlseder:2020onx}. For each of these three classes of telescopes we have estimated the largest attainable effective area of the full system of telescopes.

\section{Results}
\label{sec:results}

\begin{figure}[h!]
\begin{center}
\includegraphics[width=.6\columnwidth]{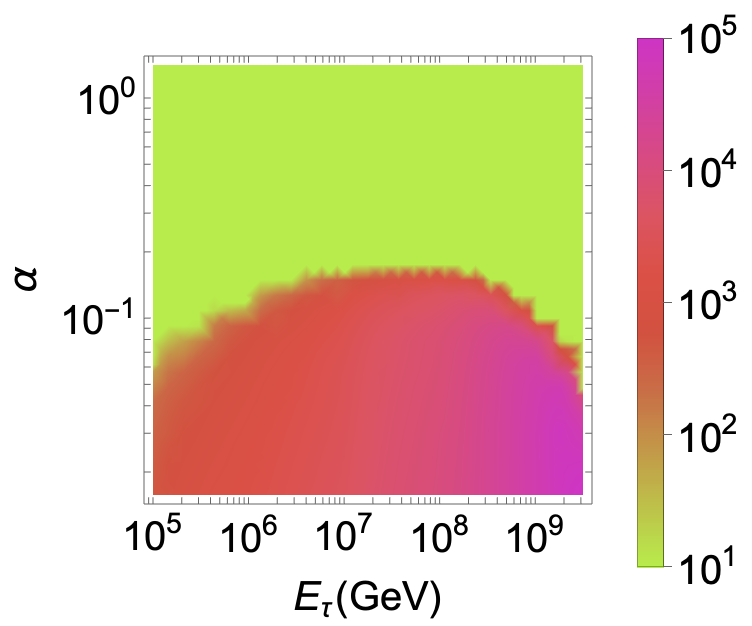}\\
\includegraphics[width=.6\columnwidth]{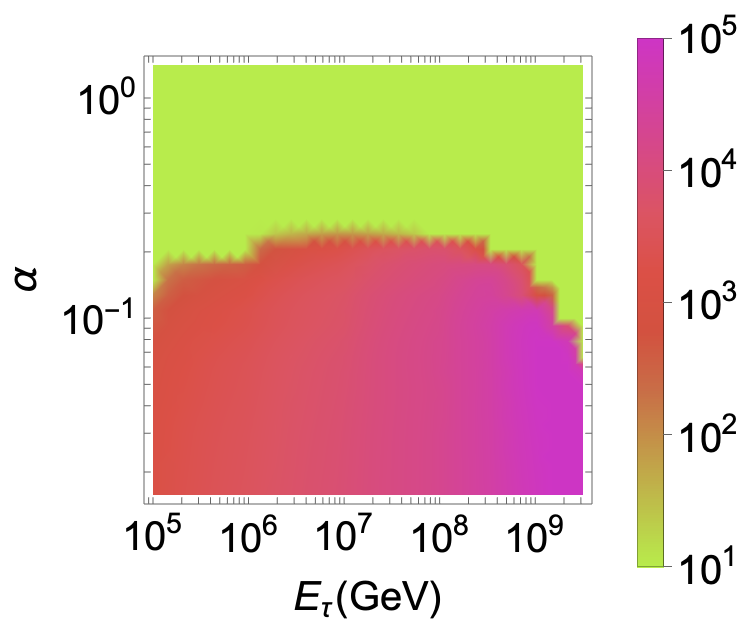}\\
\includegraphics[width=.6\columnwidth]{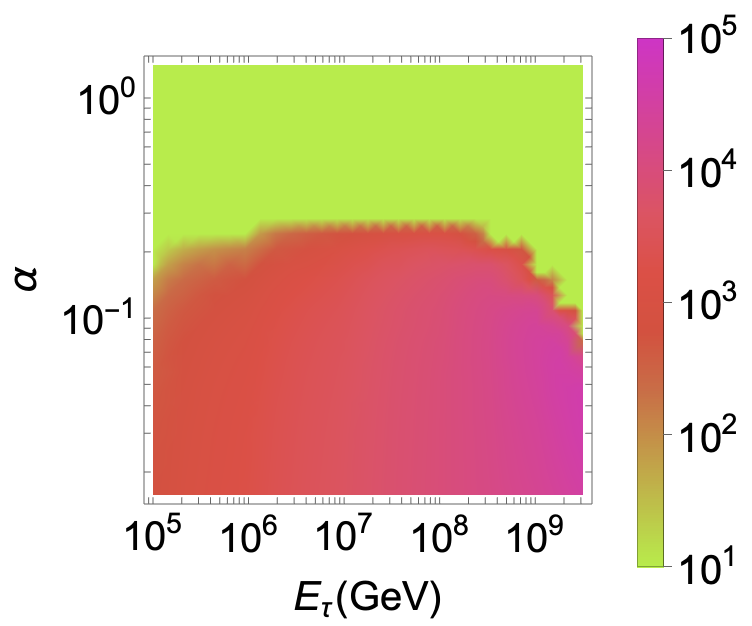}
\end{center}
\caption{Tau effective area in km$^2$ as a function of $\alpha$ and $E_\tau$ for CTA-SST, MST and LST, from top to bottom respectively.}\label{fig:taueffarea}
\end{figure}
The aim of this work is the determination of the effective area for tau neutrinos of the CTA telescope. A necessary intermediate step, which is interesting on its own, is the evaluation of $\tau$ effective area, which has been defined above. We have determined the effective area as a function of the energy and the direction of the tau-lepton. For ease of presentation, however, we have integrated this effective area over the azimuthal angle:
\begin{equation}
\overline{A}=\int_0^{2\pi} A_{eff}^\tau \, d\phi\,\,\,\,.
\end{equation}
For an isotropic flux, in fact, the distribution of tau leptons exiting the Earth is independent of $\phi$, hence we do not lose any information by performing the above integration. It is worthwhile noticing that due to a mild dependence of the previous expression on the angle $\beta$, the computation has been performed for $\beta=0$. The tau effective areas  in km$^2$ for the three configuration of the CTA, SST, MST and LST, are shown in Figure \ref{fig:taueffarea} as density plots in the $E_\tau$-$\alpha$ plane, with $\alpha$ being the exiting angle of the tau lepton with respect to the horizontal.

\begin{figure}[h!]
\begin{center}
\includegraphics[width=.6\columnwidth]{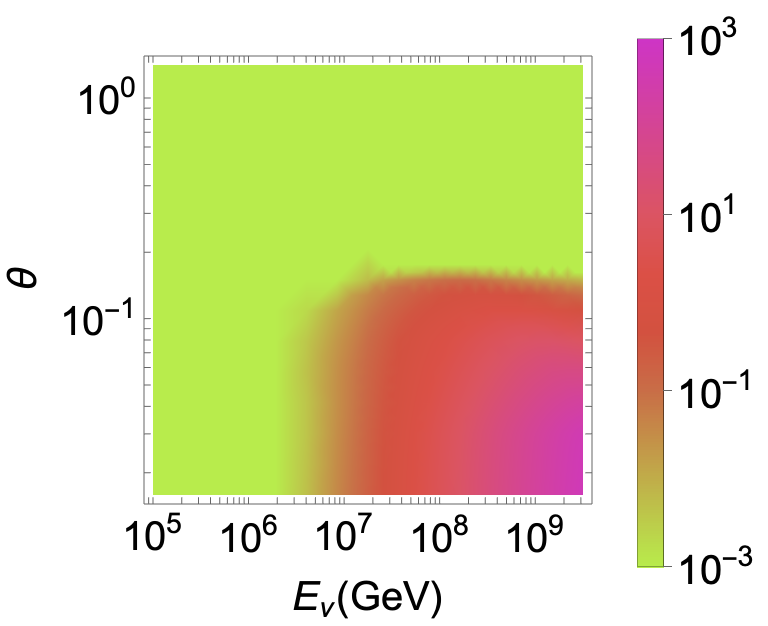}\\
\includegraphics[width=.6\columnwidth]{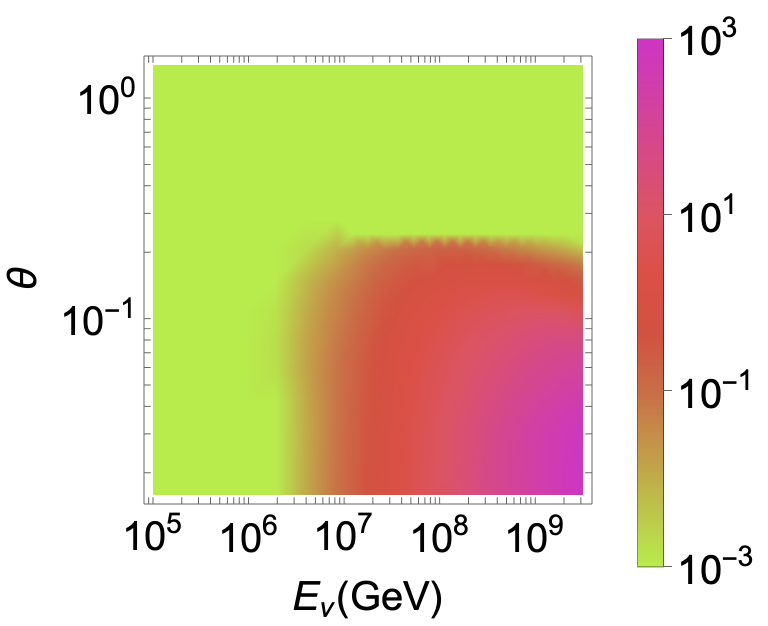}\\
\includegraphics[width=.6\columnwidth]{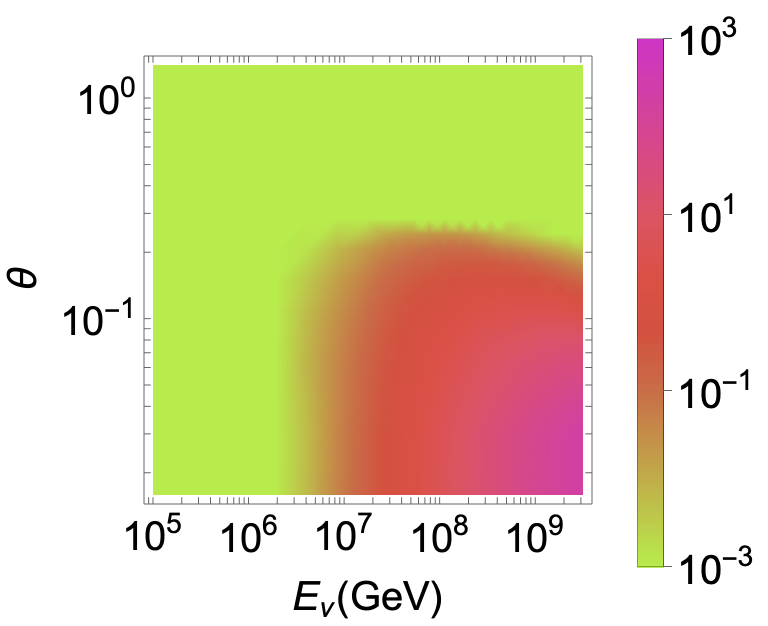}
\end{center}
\caption{Tau neutrino effective area in km$^2$ as a function of $\theta$ and $E_\nu$ for CTA-SST, MST and LST, from top to bottom respectively.}\label{fig:neueffarea}
\end{figure}

Using the procedure outlined above, the neutrino effective area can be obtained by integrating over the kernel of propagation of neutrinos through the Earth. The results for the neutrino effective area are shown in Figure \ref{fig:neueffarea}, as density plots in the $E_\nu$-$\theta$ plane, $\theta$ now being the angle of the neutrino entering the Earth with respect to the plane tangent to the Earth at the detector. From top to bottom the plots correspond to CTA-SST, MST and LST.

We checked the order of magnitude of our results against previous calculations in the literature for the MAGIC telescopes: in particular, we compared the aperture, defined as the effective area integrated over the full solid angle, for a single MST telescope with the results of Ref. \cite{Gora:2017pre}. We found a reasonable agreement in order of magnitude, confirming the validity of our method.

\begin{figure}[h!]
\begin{center}
\includegraphics[width=.9\columnwidth]{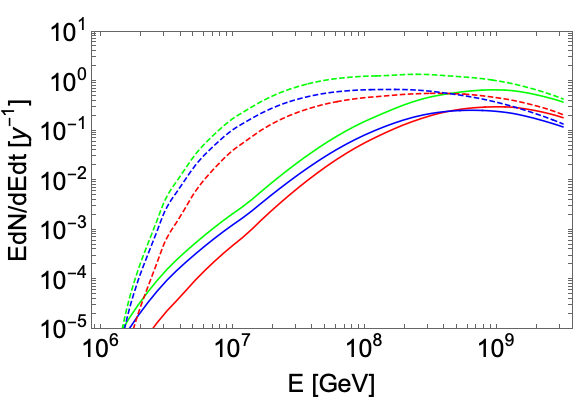}\caption{Differential number of events per unit energy per unit time expected at CTA for the SST (red), MST (green) and LST (blue) configurations: the solid lines correspond to the cosmogenic flux, the dashed lines to the FSRQ flux.}\label{fig:counting}
\end{center}
\end{figure}

Finally, it is of interest to have an idea of how many events are expected to be seen at CTA for a model flux in the energy range of interest. In the ultrahigh energy region it has been long expected that a dominant source of neutrinos is the Greisen-Kuzmin-Zatsepin (GZK) process, which gives rise to the so called cosmogenic neutrinos \textit{via} the photohadronic interaction of cosmic protons with the Cosmic Microwave Background Photons (CMB). A typical estimate of the cosmogenic spectrum is provided in Ref. \cite{Ahlers:2012rz}: in particular, we have used in this work the cosmogenic spectrum obtained there under the assumption of pure proton cosmic rays and a cosmological evolution proportional to the Star Formation Rate. \\ On the other hand, it has been recently suggested that larger neutrino fluxes might be produced by Flat Spectrum Radio Quasars: the corresponding diffuse flux has been estimated in Ref. \cite{Righi:2020ufi}. \\ For both these neutrino sources, we have estimated the differential number of events per unit energy expected to be detected by CTA, for the cases of SST, MST and LST. The results are shown in Figure \ref{fig:counting}.

\section{Conclusions}
\label{sec:conclusion}

The Cherenkov Telescope Array (CTA) belongs to the next generation of IACTs, and it is going to certainly represent a breakthrough in the detection capability of high-energy cosmic rays. In the mean time, the Neutrino Astronomy has just become a reality with apparatus like km$^3$-Neutrino Telescopes (IceCube, Antares and  Km3Net) and with their first observation of astrophysics high-energy neutrinos. In past it has been discussed in literature the possibility to use cosmic rays detectors to detect ultra-high-energy neutrinos just by looking at nearly horizontal events or even Earth-skimming ones, see for example the Pierre Auger Observatory limit \cite{Abraham:2009uy}. It is worth while reminding that Earth-skimming events can be induced by $\nu_\tau$ only and hence represent a way to measure the flavour of the arriving neutrino flux. In our analysis we have scrutinised the possibility to use CTA to detect Earth-skimming tau neutrinos. As shown in the previous section, the predictions seem promising. In the optimistic case of larger neutrino fluxes  produced by Flat Spectrum Radio Quasars for example, one expects few event of $\nu_\tau$-skimming per year for energies above $10^8$ GeV. The prediction is almost reduced by an order of magnitude for more conservative cosmogenic neutrino fluxes, but also in this case few events would be collected in a decade of running.  Thus, our estimate seems to suggest that apparatus like CTA could also perform as Neutrino Telescope at least in the Ultra-High-Energy range.
\\ \\
{\bf Acknowledgments}: This work was partially supported by the research grant number 2017W4HA7S "NAT-NET: Neutrino and Astroparticle Theory Net- work" under the program PRIN 2017 funded by the Italian Ministero dell'Universit\`a e della Ricerca (MUR). The authors acknowledge partial support by the research project TAsP (Theoretical Astroparticle Physics) funded by the Instituto Nazionale di Fisica Nucleare (INFN).

\appendix 

\section{Cherenkov light produced in an extensive air shower}
\label{app1}

The production of light from extensive air showers can be regarded as the superposition of the Cherenkov light produced by each particle in the shower. A single charged particle produces a number of photons $dQ/ds$ per unit path length
\begin{equation} \label{spectrum}
\frac{dQ}{ds}=\int d\omega \frac{d\zeta}{2\pi} \frac{q_0}{c} \sin^2 \theta \simeq \int d\omega \frac{d\zeta}{2\pi} \frac{q_0}{c} \theta^2,
\end{equation}
where $q_0=1/137$, $\omega$ is the light frequency, $\zeta$ is the azimuth angle around the direction of the charged particle and $\theta$ is the Cherenkov angle, which has been assumed to be very small. This is usually the case for extensive air showers. The Cherenkov angle is connected to the energy by the relation
\begin{equation}
\cos\theta=\frac{c}{n(\omega) v}.
\end{equation}
The refraction index is generally slowly varying with frequency in the optical range, so we can take it to be constant, with the integration in \eqref{spectrum} being performed only over the range of interest for the detector. In our case this range is between the wavelengths of $\lambda_1=300$ nm and $\lambda_2=600$ nm. If we integrate over this wavelength range, after setting
\begin{equation}
Q_0=q_0 \frac{2\pi(\lambda_2-\lambda_1)}{\lambda_1\lambda_2},
\end{equation}
we find
\begin{equation}
\frac{dQ}{ds}=\frac{d\zeta}{2\pi} \, Q_0 \, \theta^2.
\end{equation}

In the following, we will mostly neglect such details as the dependence on the altitude of the refraction index or the absorption coefficients. Since, however, as we will see, most of the Cherenkov light production comes from a single point of the shower, it is enough to insert these factors in the final formulas evaluated at such maximum production point.

There is also need for a parameterization of the differential energy spectrum of the shower. Detailed such parameterizations can be found in Ref. \cite{deNaurois:2009ud}. 

Finally, for ease of mathematical treatments, the angular distribution of the charged particles in the shower has been taken from Ref. \cite{sokolsky2018introduction}, where it is given in the form coming from Moliere theory as:
\begin{equation}
e^{-\phi \alpha^2} 2\phi \alpha d\alpha,
\end{equation}
where $\alpha$ is the angle of the particle and $\phi=A E^2 + B E$.

The quantity which is of greatest experimental interest is the number of photons per unit area arriving on a surface at a distance $S$ from the origin of the shower. If a photon is produced at a distance $s$ from the shower core, from a charged particle with an angle $\alpha$ from the direction of the primary particle, it will travel along a cone of angular width $\theta$, the Cherenkov angle, which can be parameterized in terms of the distance traveled $\lambda$ and the azimuth around the cone axis $\zeta$ as
\begin{eqnarray}
&&\mathbf{r}(\lambda,\zeta)=(0,0,s)+\lambda\left[\cos\theta (\sin\alpha,0,\cos\alpha) \right.\nonumber\\
&+&\left.\sin\theta\cos\zeta(0,1,0)\right.\nonumber \\ &+& \left. \sin\theta\sin\zeta (\cos\alpha,0,-\sin\alpha)\right],
\end{eqnarray}
where we have chosen the $z$ axis along the shower direction.

From this expression it is easy to obtain the lateral distance of the photon when it arrives the plane at $z=S$, which, in the case of small $\alpha$ and $\theta$, is $(S-s)(\alpha^2+\theta^2+2\alpha\theta\sin\zeta)$.

We can now express the number of photons per unit area arriving at the detector as
\begin{eqnarray}
\frac{dQ}{dA}&=&\frac{1}{\pi}\int ds \, d\alpha \, dE \, \frac{d\zeta}{2\pi} Q_0 \theta^2 \frac{dN}{dE}(E,s) e^{-\phi \alpha^2} 2\alpha \phi \nonumber\\
&\times& \delta\left(r^2-(S-s)^2 (\alpha^2+\theta^2+2\alpha\theta\sin\zeta)\right).
\end{eqnarray}
The integral over $\zeta$ can be used to eliminate the delta function. After redefining $\alpha^2 \phi=\mu$ we find
\begin{eqnarray}
\frac{dQ}{dA}&=&\frac{Q_0}{2\pi^2} \int \frac{ds}{(S-s)^2} \frac{dN}{dE} dE \theta^2 \nonumber\\
&\times &\int_{0}^{+\infty} \frac{e^{-\mu}d\mu}{\sqrt{-\frac{\mu^2}{\phi^2}+2\frac{\mu}{\phi}(\theta^2+\chi^2)-(\chi^2-\theta^2)^2}},
\end{eqnarray}
where we have defined $\chi\equiv r/(S-s)$.

After making the transformation $\mu\equiv\phi(\theta^2+\chi^2)+2\theta\chi\phi\sin\tau$ the integral over $\mu$ is recognized to be a modified Bessel function of the first kind and zero order, so that we are led to the final result
\begin{equation}
\frac{dQ}{dA}=\frac{Q_0}{2\pi}\int \frac{ds}{(S-s)^2} \frac{dN}{dE} dE \theta^2 \phi e^{-\phi(\theta^2+\chi^2)} I_0(2\theta \chi \phi).
\end{equation}
It has to be kept in mind here that $\theta$ depends implicitly on the energy.

\section{Geometry of the simulation}
\label{app2}

The first geometrical constraint considered in the simulation correspond to the fact that the $\tau$ decay point, $D$, is before the plane $\pi$ (see Figure \ref{geometry1}). This corresponds to the fact that the distance $S$ has to be positive, that is
\begin{equation}
h\,\sin\alpha-\sigma\cos\alpha\cos (\gamma-\phi) > 0.
\end{equation}
Note that the previous inequality expresses the condition that the scalar product between the vector $P'-P$ and the direction $\mathbf{m}$ is positive.

Then, we have to impose that the maximum of the shower, $D$, is not beyond the plane $\pi$. In order to write down this condition, let us consider the point $M=P+[l_\tau(E_\tau) + s_{max}]\,\mathbf{n}$, where $s_{max}$ represents the distance of the maximum of the shower from the decay point of the $\tau$, that is the distance between the points $D$ and $M$. We need to impose that the scalar product between $M-P'$ and $\mathbf{m}$ is positive, that is
\begin{eqnarray}
&&\sigma\sin\gamma\cos\beta + [l_\tau(E_\tau) + s_{max}] (\cos\beta\cos\alpha\sin\phi-\nonumber\\
&& ~~~~~\sin\beta\sin\alpha) + h\,\sin\beta>0.
\end{eqnarray}

We now discuss how to correct the previous relations to take into account the Earth curvature. We parameterize the Earth surface by two angles, $\theta$ and $\gamma$, so that the coordinates of a point on the sphere are $(R\sin\theta\cos\gamma,R\sin\theta\sin\gamma,R\cos\theta)$. In terms of the distance $\sigma$ along the Earth from the pole, where we locate the telescope, $\theta=\sigma/R$. Let the tau lepton exit the Earth at the coordinates $\sigma$ and $\gamma$. The direction of the tau lepton will be inclined at angle $\alpha$ with respect to the normal to the surface at the exiting point. The normal direction is given, as usual, by $\mathbf{n}=(\sin\theta\cos\gamma,\sin\theta\sin\gamma,\cos\theta)$. The two tangent vectors to the surface can be chosen as $\mathbf{t}_1=(\sin\gamma,-\cos\gamma,0)$ and $\mathbf{t}_2=(\cos\theta\cos\gamma,\cos\theta\sin\gamma,-\sin\theta)$. Therefore the exiting direction of the lepton is
\begin{equation}
\mathbf{m}=\sin\alpha\,\mathbf{n}+\cos\alpha\cos\delta\,\mathbf{t}_1 + \cos\alpha\sin\delta\,\mathbf{t}_2.
\end{equation}
In the previous sections we parameterized the exiting direction of the tau lepton by its angle with the horizontal plane, $\alpha$, and the azimuthal angle, $\phi$, that is the angle that the projected exiting direction on the Earth surface forms with the $x$ axis. The most natural extension of this definition is given by the requirement that the ratio between the $y$ and the $x$ component of the projection of $\mathbf{m}$ on the Earth plane is equal to $\tan\phi$,
\begin{equation}
\frac{\sin\delta\sin\gamma-\cos\delta\cos\theta\cos\gamma}{\cos\delta\cos\theta\sin\gamma+\sin\delta\cos\gamma}=\tan\phi.
\end{equation}
This can be solved to give
\begin{equation}
\tan\delta=\cos\theta\frac{\cos\left(\gamma-\phi\right)}{\sin\left(\gamma-\phi\right)}.
\end{equation}
Having fixed $\delta$, the direction $\mathbf{m}$ is completely determined, so that the value of $S$ is determined by the condition
\begin{equation}
S=\mathbf{m}\cdot(P'-P),
\end{equation}
where $P'=(0,0,R+h)$ is the detector position and $P$ is the exiting point of the tau lepton. This can be rearranged in the form
\begin{eqnarray}
S=-R\sin\alpha\sin^2\theta-R\cos\alpha\sin\delta\sin\theta\cos\theta\nonumber\\
+(R+h-R\cos\theta)(\sin\alpha\cos\theta-\cos\alpha\sin\delta\sin\theta).
\end{eqnarray}

\end{document}